\documentclass[showpacs,twocolumn,aps,floatfix,superscriptaddress]{revtex4}
\usepackage{amsmath,amssymb,eucal,graphicx,subfigure}
\begin{document}
\title{Front Propagation with Rejuvenation in Flipping Processes}
\author{T.~Antal}
\affiliation{Program for Evolutionary Dynamics, Harvard University,
Cambridge, Massachusetts, 02138 USA}
\author{D.~ben-Avraham}
\affiliation{Physics Department, Clarkson University, Potsdam, New York
13699 USA} 
\author{E.~Ben-Naim}
\affiliation{Theoretical Division and Center for Nonlinear
Studies, Los Alamos National Laboratory, Los Alamos, New Mexico
87545 USA}
\author{P.~L.~Krapivsky}
\affiliation{Theoretical Division and Center for Nonlinear
Studies, Los Alamos National Laboratory, Los Alamos, New Mexico
87545 USA}
\affiliation{Department of Physics,
Boston University, Boston, Massachusetts 02215 USA}
\begin{abstract}
We study a directed flipping process that underlies the performance of
the random edge simplex algorithm. In this stochastic process, which
takes place on a one-dimensional lattice whose sites may be either
occupied or vacant, occupied sites become vacant at a constant rate
and simultaneously cause all sites to the right to change their state.
This random process exhibits rich phenomenology. First, there is a
front, defined by the position of the left-most occupied site, that
propagates at a nontrivial velocity. Second, the front involves a
depletion zone with an excess of vacant sites. The total excess
$\Delta_k$ increases logarithmically, $\Delta_k \simeq \ln k$, with
the distance $k$ from the front. Third, the front exhibits
rejuvenation --- young fronts are vigorous but old fronts are
sluggish. We investigate these phenomena using a quasi-static
approximation, direct solutions of small systems, and numerical
simulations.
\end{abstract}
\maketitle

\section{Introduction}

\begin{figure}[h]
\includegraphics[width=0.2\textwidth]{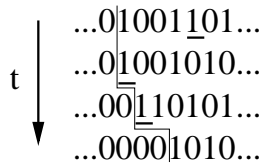}
\caption{Illustration of the flipping process. The arrow indicates the
direction of time and the line indicates the position of the advancing
front.}
\label{fig-flip}
\end{figure}

The simplex algorithm \cite{gbd} is the fastest general algorithm
for solving linear problems. While efficient in the typical case, the
deterministic simplex algorithm requires an exponential time in the worst
cases \cite{km,kk}. Randomized versions of the simplex algorithm
have an improved running time that is quadratic in the number of
inequalities. The performance of the random edge simplex algorithm on
Klee-Minty cubes \cite{km} ultimately reduces to a simple asymmetric
flipping process in one dimension \cite{ghz}. In this process, an
infinite sequence of 0 and 1 bits evolves by flipping randomly chosen
1 bits and simultaneously flipping all bits to the right. Figure
\ref{fig-flip} illustrates how the underlined bit flips all bits to
the right. When flips occur at a constant and spatially uniform rate,
the position of the left-most 1 bit moves to the right at a constant
average velocity. Previous formal studies were primarily concerned
with establishing the ballistic front motion rigorously \cite{pem},
yet most of the questions concerning the flipping process, including
the propagation velocity, remain largely unanswered.

We approach this random process as a nonequilibrium dynamics problem
and by utilizing a host of theoretical and computational methods, we
find that this directed flipping process exhibits interesting
phenomenology beyond the ballistic front propagation. We also
propose a modified process where front propagation is forbidden and
show that this process, for which further theoretical analysis is
possible, provides an excellent quantitative description.

Our starting point is a quasi-static approximation. In this
description, the shape of the propagating front is assumed to be fixed
and additionally, spatial correlations are ignored. This approximation
yields a qualitative description for the overall shape of the front
and an exact description for the shape far away from the front. The
propagating front consists of a depletion zone as the number of 0 bits
exceeds the number of 1 bits, and the cumulative depletion grows
logarithmically with distance from the front.

Direct numerical simulations of the flipping process reveal that
spatial and temporal correlations are substantial. In general,
neighboring bits are correlated as manifested by the increased
likelihood of finding consecutive strings of identical bits. There
are also aging and rejuvenation. The state of the system strongly
depends on age, defined as the time elapsed since the most recent
front advancement event. In particular, young fronts are more rapid
than old front.

We also develop a formal solution method that describes the
evolution of a finite segment that includes the front. In this
approach, the time evolution of all microscopic configurations of a
finite segment is described under the assumption that the system is
completely random outside the segment. The predictions improve
systematically as the segment size increases but there is a
limitation since the number of configurations grows exponentially
with segment size. Nevertheless, we are able to obtain accurate
estimates for quantities of interest including the propagation
velocity by using Shanks extrapolation.

In the directed flipping process, the system does not reach a steady
state because of the perpetual motion of the front, yet when the front
is pinned, the system does settle into a steady state. We therefore
also examined a modified process in which the flipping of the leftmost
bit is forbidden. Remarkably, this pinned front process provides an
excellent quantitative approximation of the original propagating front
process. In this case, we are able to obtain several exact
results. For example, we can show that a pair of neighboring sites is
correlated. Moreover, the small system solution is now exact and
combined with the Shanks transformation, yields excellent results for
the velocity.

The rest of this paper is organized as follows. In section II,
titled ``propagating fronts'', we investigate the original flipping
process. We begin with a quasi-static approximation for the shape of
the front, continue with numerical simulations that elucidate
spatial and temporal correlations, and finish with analysis of small
segments. In section III, titled ``pinned fronts'', we examine the
corresponding behaviors in a modified flipping process where the
front is pinned and hence further theoretical analysis is possible.
Conclusions are presented in section IV.

\section{Propagating Fronts}

The flipping process takes place on an infinite one-dimensional
lattice whose sites may be in one of two states. If $\sigma_i$ denotes
the state of $i$th site then $\sigma_i=1$ corresponds to an occupied
site, a 1-bit, and $\sigma_i=0$ corresponds to a vacant site, a
0-bit. In the flipping process, each occupied site may ``flip'' from
the occupied state to the vacant state and consequently cause all
sites to the right to simultaneously change their state. For example,
when the $j$th site flips,
\begin{equation}
\label{process}
\sigma_i\to 1-\sigma_i, \quad{\rm for\ all}\quad i\geq j.
\end{equation}
The flipping process is uniform: all occupied sites flip at a uniform
rate, set to one without loss of generality. Note that the interaction
range is infinite: every flip event affects an infinite number of
sites! This is in contrast, for example, with constrained spin
dynamics such as the east model \cite{rs,se,ad} where the flipping is
caused only by the neighboring spin on the left.

Vacant sites with no occupied sites to their left remain vacant
forever.  Moreover, the left-most occupied site defines a front that
advances to the right, as shown in figure 1. We consider the natural
initial condition where all sites left of the origin are vacant,
$\sigma_i(t=0)=0$ for all $i<0$, the origin is occupied
$\sigma_0(t=0)=1$, and all sites right of the origin are randomly
occupied: with equal probabilities $\sigma_i(t=0)=1$ or
$\sigma_i(t=0)=0$ for all $i>0$.

\subsection{Front Profile and Depletion}  

We index the system using a reference frame that is moving with the
front. Specifically, we characterize lattice sites by their distance
$k$ from the front, and by definition, $\sigma_0=1$. The profile of
the advancing front is best described by the density $\rho_k(t)$, the
average occupation at distance $k$ from the front at time $t$,
$\rho_k(t)\equiv\langle \sigma_k(t)\rangle$, where the brackets
indicate an average over all realizations of the random process.

Our theoretical description involves two simplifying assumptions. If
we overlook the motion of the front, the densities satisfy
\begin{eqnarray}
\label{nav-eq}
\frac{d\,\langle \sigma_k\rangle}{dt}\!=\!
\left\langle\!\Bigg(\sum_{j=0}^{k-1}\sigma_j\Bigg)\!(1-\sigma_k)\!\right\rangle
\!-\!\left\langle\!\Bigg(1+\sum_{j=0}^{k-1}
\sigma_j\Bigg)\sigma_k\!\right\rangle\!
\end{eqnarray}
for $k>0$. The gain term on the right-hand size accounts for vacant
sites changing into occupied sites and conversely, the loss term
represents occupied sites changing into vacant sites.  Since every
occupied site to the left can cause a vacant site to change, the
gain rate at the $k$th site equals the total number of occupied
sites to the left.  The loss rate, however, is larger by one because
a flip at the site itself can also cause an occupied site to change.

The evolution equations \eqref{nav-eq} are hierarchical: the
equation for one-site averages involves two-sites averages, the
equation for two-site averages involves three-site averages, etc. If
we ignore possible correlations between different sites and
approximate two-site averages by the product of the respective
single site averages $\langle \sigma_j\sigma_k\rangle\to \langle
\sigma_j\rangle\langle \sigma_k\rangle$, the densities satisfy the
closed equation
\begin{equation}
\label{rho-eq}
\frac{d\rho_k}{dt}=
\left(\sum_{j=0}^{k-1}\rho_j\right)(1-\rho_k)
-\left(1+\sum_{j=0}^{k-1}\rho_j\right)\rho_k.
\end{equation}
The flipping rates in equations \eqref{nav-eq}-\eqref{rho-eq}
reflect the fact that occupied sites change at a higher rate than
vacant sites.

\begin{figure}[t]
\includegraphics[width=0.4\textwidth]{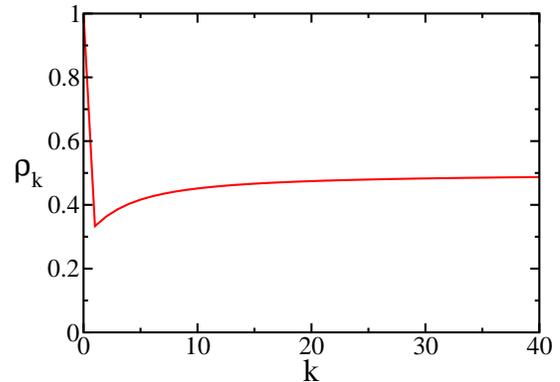}
\caption{The density profile $\rho_k$, obtained from the
quasi-static approximation.}
\label{fig-ck}
\end{figure}

Our final assumption is that in the reference frame moving with the
front, the system is quasi-static. Indeed, by definition
$d\rho_0/dt=0$, and we further assume \hbox{$d\rho_k/dt=0$} for all
$k$.  The stationary density profile is
\begin{equation}
\label{rhok-sol}
\rho_k=\frac{\sum_{j=0}^{k-1}\rho_j}{2\sum_{j=0}^{k-1}\rho_j+1}.
\end{equation}
This recursive equation is solved subject to the boundary condition
$\rho_0=1$.  For small $k$ we have
\begin{equation}
\label{rho_front}
\rho_k=1,\frac{1}{3},\frac{4}{11},\frac{56}{145},\qquad k=0,1,2,3,\cdots.
\end{equation}

\begin{figure}[t]
\includegraphics[width=0.45\textwidth]{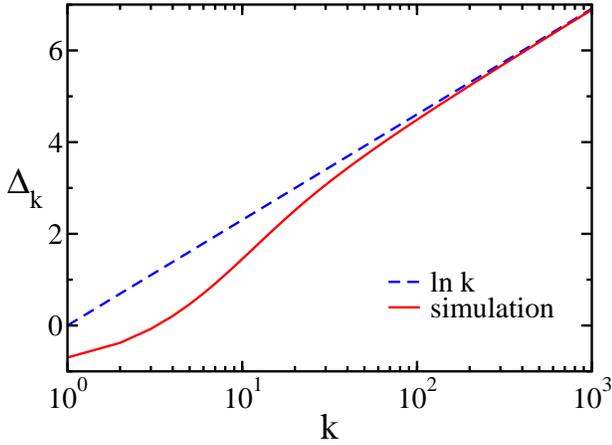}
\caption{The total excess of empty sites $\Delta_k$ versus distance
$k$.  The simulations are in a system of size $L=1000$.}
\end{figure}

Despite the crude simplifying assumptions, this quasi-static
approximation provides the following valuable insights (see figure
\ref{fig-ck}):
\begin{enumerate}
\item{\em Depletion.} With the exception of the occupied front, all
sites are more likely to be vacant, $\rho_k<1/2$ for all $k>0$. In
other words, the propagating front includes a depletion zone. This
depletion is a direct consequence of the fact that occupied sites
change at a higher rate than vacant sites. In other words, vacant
sites have a larger lifetime.
 \item {\em Monotonicity}.  The density profile is monotonic,
$\rho_i>\rho_j$ for $i>j\ge 1$.
\end{enumerate}

The tail of the density profile can be obtained by noting that
$\rho_k\to1/2$ as follows from \eqref{rhok-sol}. Consequently, the
average total ``mass'' to the left of a given site,
\hbox{$m_k=\sum_{j=0}^{k-1}\rho_j$}, grows linearly with distance,
$m_k\simeq k/2$. At large distances, the recursion equation for the
density $\rho_k=m_k/(2m_k+1)$ can be re-written as
$\rho_k\simeq\frac{1}{2}-\frac{1}{4m_k}$ and therefore,
\begin{equation}
\label{rhok-tail}
\rho_k\simeq \frac{1}{2}-\frac{1}{2k}.
\end{equation}
Far away from the front, sites are occupied at random as $\rho_k\to
1/2$ for $k\to\infty$. Indeed, sites at the tail change their state
extremely rapidly at rates that grow linearly with distance. These
rapid changes effectively destroy spatial correlations. Moreover,
the advancement of the front becomes irrelevant at large distances.
Hence, the two assumptions underlying our theory are inconsequential
in the tail region and \eqref{rhok-tail} is in fact exact. We
comment that the algebraic tail \eqref{rhok-tail} is unusual because
traveling waves are typically characterized by exponential tails
\cite{wvs,bd}.

The cumulative expected excess of vacant sites over occupied sites,
$\Delta_k=\sum_{j=0}^{k-1} (1-2\rho_j)$, measures the extent of the
depletion zone. This quantity follows from the tail behavior
\eqref{rhok-tail}, $\Delta_k=k-2m_k$, and since \hbox{$m_k\simeq(k-\ln
k)/2$}, the excess of vacant sites grows logarithmically with
distance,
\begin{equation}
\label{excess}
\Delta_k\simeq \ln k.
\end{equation}
Thus, the total excess of vacant sites is divergent!

We confirmed the theoretical predictions for the algebraic tail
\eqref{rhok-tail} and the logarithmic growth of the excess
\eqref{excess} using massive Monte Carlo simulations (figure 2). The
numerical simulations are straightforward. In each simulation step one
site is chosen at random.  If this site is occupied, the state of the
site and all sites to the right change according to \eqref{process},
but otherwise, nothing happens.  After each step, time is augmented by
the inverse of the system size $t\to t+L^{-1}$ where $L$ is the number
of sites in the lattice. In our implementation, the front is always
located at the zeroth site, $\sigma_0=1$. Whenever the front advances
by $n$ sites, all lattice sites are appropriately shifted to the left,
$\sigma_i\to \sigma_{i-n}$ (the $n$ rightmost sites are reoccupied at
random). Subsequently, the front position is augmented by $n$. This
efficient implementation allows us to simulate the evolution of the
system up to extremely large times. We can evolve a system of size
$L=10^2$ up to time $t=10^{11}$, and we obtain statistical averages
from snapshots of the system taken at unit time intervals.

\begin{figure}[t]
\includegraphics[width=0.45\textwidth]{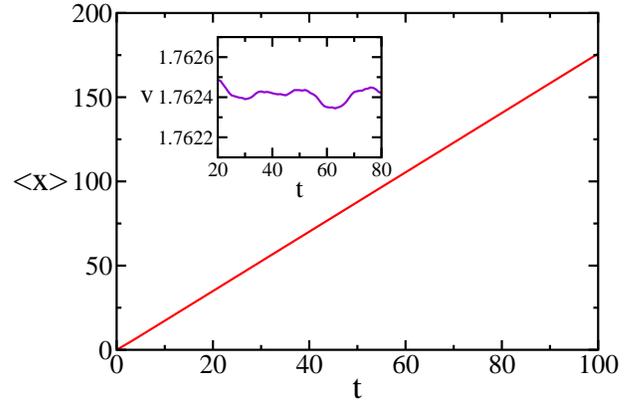}
\caption{The average position of the leftmost bit $\langle x\rangle$
versus time $t$. The results are from a Monte Carlo simulation in a
system of size $L=10^2$, evolved up to time $t=10^{11}$. The inset
shows the velocity $v=d\langle x\rangle/dt$ versus time.}
\label{fig-xt}
\end{figure}

\subsection{Front Propagation} 
Whenever the leftmost site flips, the front position $x$ advances by
$n$ lattice sites,
\begin{equation}
\label{string} \cdots0000\underbrace{11111}_n0100\cdots \to
\cdots0000\underbrace{00000}_n1011\cdots.
\end{equation}
Hence, the leftmost string of occupied sites governs the front
propagation. Like all other sites, the front flips at a unit rate,
and consequently, the average front position grows ballistically,
\begin{equation}
\langle x\rangle \simeq v\,t,
\end{equation}
and the propagation velocity $v$ equals the average size of the
leftmost occupied string $v=\langle n\rangle$.

Let $S_n$ be the probability that the leftmost $n$ lattice sites
including the front are all occupied,
\begin{equation}
\label{sn} S_n\equiv {\rm Prob}(\underbrace{11111}_n).
\end{equation}
The probability
of finding a string of exact length $n$ as in \eqref{string} is equal
$S_n-S_{n+1}$, and therefore the velocity is given by \hbox{$v=\langle
n\rangle=\sum_{n=1}^{\infty} n(S_n-S_{n+1})$}. Consequently, 
the velocity equals the sum of string
probabilities
\begin{equation}
\label{v-eq} v=\sum_{n=1}^\infty S_n.
\end{equation}

In the Quasi-Static Approximation (QSA), correlations between
different sites are neglected, and hence, the string probability
\eqref{sn} is a product over the corresponding densities,
\begin{equation}
\label{pn-qsa}
S^{\rm uncorr}_n=\rho_1\rho_2\cdots\rho_{n-1},
\end{equation}
for $n>1$ while $S_1=1$.  With this approximate expression, the
propagation velocity is
$v=1+\rho_1+\rho_1\rho_2+\rho_1\rho_2\rho_3+\cdots$.  We obtain the
approximate velocity
\begin{equation}
\label{v-qsa}
v_{\rm QSA}=1.534070
\end{equation}
by substituting the densities from \eqref{rhok-sol} into
\eqref{pn-qsa} and then summing numerically.  The velocity
\eqref{v-qsa} obeys the obvious bounds $1\leq v\leq 2$. The lower
bound reflects that the front must advance by at least one lattice
site, and the upper bound corresponds to the completely random
configuration, $\rho_k=1/2$ and $S_n=2^{-(n-1)}$.

\begin{figure}[t]
\includegraphics[width=0.45\textwidth]{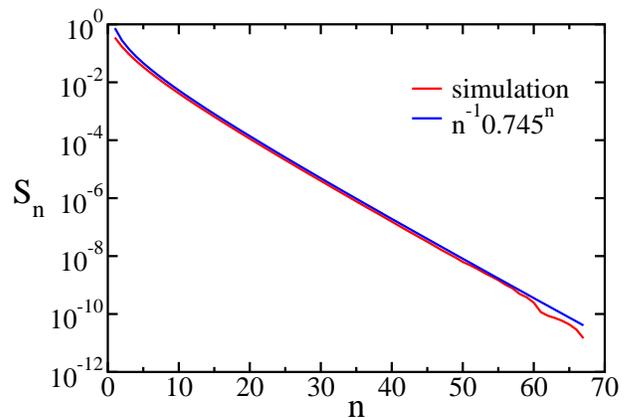}
\caption{The string probability $S_n$ versus the string length $n$.
The results are from a Monte Carlo simulation in a system of size
$L=200$, evolved up to time $t=10^{11}$.}
\label{fig-sn}
\end{figure}

The numerical simulations confirm that the front advances
ballistically (figure \ref{fig-xt}) but the propagation velocity is
larger than the value predicted by the quasi-static approximation
\begin{equation}
\label{v-mc} v_{\rm MC}=1.7624\pm 0.0001.
\end{equation}
Strong spatial correlations are primarily responsible for the
discrepancy between \eqref{v-qsa} and \eqref{v-mc}. Indeed, if we
substitute the densities $\rho_k$ obtained from the Monte Carlo
simulations into the product expression \eqref{pn-qsa} and perform
the summation in \eqref{v-eq}, we obtain the value $v=1.5329\pm
0.0001$ that is surprisingly close to the quasi-static approximation
\eqref{v-qsa}. We therefore conclude that spatial correlations
between neighboring sites have a significant effect on the velocity.

\subsection{Correlations, Aging, and Rejuvenation}  

Spatial structures and spatial correlations can be quantified in
multiple ways and we focus on the likelihood of occupied strings
$S_n$.  Numerically, we find that this quantity decays exponentially
(figure \ref{fig-sn}),
\begin{equation}
S_n\sim n^{-\nu}\lambda^n,
\end{equation}
as $n\to\infty$ with $\lambda=0.745\pm 0.001$ and $\nu\approx 1$.  The
quasi-static approximation yields much more rapid decay, $\lambda=1/2$
and $\nu=1$ as follows from the algebraic tail \eqref{rhok-tail} and
the product expression \eqref{pn-qsa}. Of course, when sites are
completely uncorrelated, one also has $\lambda=1/2$. The fact that
$\lambda$ is larger than $1/2$ reflects that the system is strongly
correlated. There is a significant enhancement of strings of
consecutively occupied sites and this enhancement is largely
responsible for the larger velocity \eqref{v-mc}.

Even though spatial correlations are significant and
affect quantities of interest such as the velocity, they are limited
in extent as indicated by the exponential decay of the string
likelihood. For this reason, numerical simulations may be performed
in relatively small systems.  Given the spatial extent of strings 
shown in figure \ref{fig-sn}, we performed the simulations using a
relatively small system, $L=200$. This system size is used
throughout this investigation, unless noted otherwise.

\begin{figure}[t]
\includegraphics[width=0.45\textwidth]{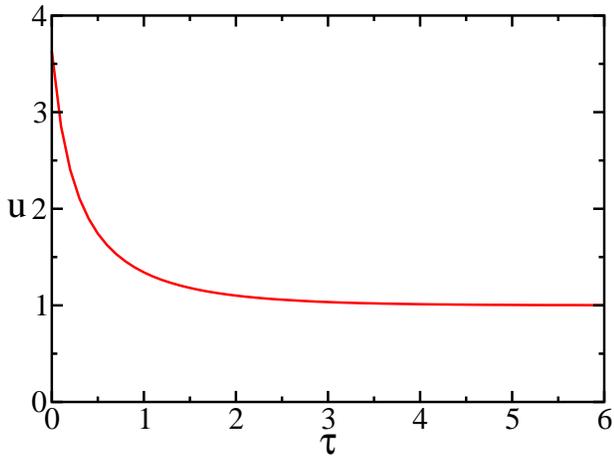}
\caption{The velocity $u$ versus age $\tau$.}
\label{fig-age}
\end{figure}

We also probed the correlation between two successive front
``jumps'' as a measure of temporal correlations. Let $n$ and $n'$ be
the sizes of two consecutive jumps, respectively. If the front
advances via a renewal process then $\langle nn'\rangle=\langle
n\rangle^2=v^2$. However, the numerical simulations yield $\langle
n\,n'\rangle=2.959\pm0.001$ while $v^2=3.1060\pm 0.0001$. Thus,
front advancement events are correlated, so the state of the system
just after a jump is correlated with the state of the system just
before a jump.

This temporal correlation affects, in particular, the diffusion
coefficient $D$ that quantifies the uncertainty in the front
position,
\begin{equation}
\label{D-def}
\langle x^2\rangle-\langle x\rangle^2 \simeq 2Dt.
\end{equation}
Numerically, we find $D=2.856\pm 0.001$. In contrast with the
velocity \eqref{v-eq} that follows from average quantities such as
the average segment density, the diffusion coefficient requires
more detailed information about temporal correlations \cite{note}.

To further characterize the dynamics, we define the age $\tau$ as
the time elapsed since the most recent front jump. Moreover, we
define the age-dependent velocity $u(\tau)$ as the average size of the
leftmost string $n$ as in \eqref{string} at age $\tau$ because this
quantity governs the front propagation. The simulations show that
the velocity rapidly decays with age (figure \ref{fig-age}). Of
course, since long-living fronts outlive any of their occupied
neighbors, $u\to 1$ as $\tau\to\infty$. Aging fronts are therefore
sluggish. In contrast, newly-born fronts are much more vigorous
because $u(0)>v$. Since the flipping process is completely random,
the survival probability of a configuration decays exponentially
with age. The average velocity in \eqref{v-eq} is the weighted
integral of the age-dependent velocity
\begin{equation}
\label{v-int} v=\int_0^\infty d\tau \,u(\tau)\,e^{-\tau},
\end{equation}
and the weight equals the exponential survival probability.

\begin{figure}[t]
\includegraphics[width=0.45\textwidth]{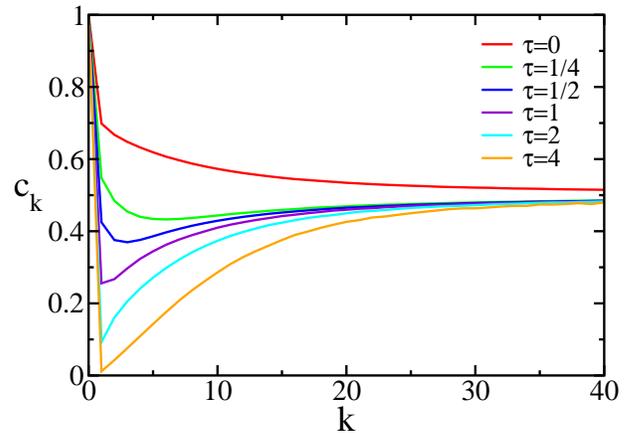}
\caption{The density profile $c_k$ versus distance $k$ at different
ages.}
\label{fig-ckt}
\end{figure}

The age-dependence of the velocity implies that the shape of the front
must also be age-dependent. We therefore measured the density profile
$c_k(\tau)=\langle \sigma_k(\tau)\rangle$, defined as the average
occupation at distance $k$ from the front at age $\tau$. We find
interesting evolution with age.  The profile of long-living fronts has
a depletion zone and is qualitatively similar to the average profile
discussed above, but the profile of newly born fronts has an
enhancement of occupied sites over vacant sites (figure
\ref{fig-ckt}). This rejuvenation is intuitive: the state of the
system just after a jump is a mirror image of the state of the system
just before the jump. Long living fronts are followed by a large
string of vacant sites, and these fronts are necessarily slow.  Yet,
upon flipping, such sluggish fronts rejuvenate as the string of vacant
sites become a string of occupied sites. Interestingly, the density
profile may even be non-monotonic at intermediate ages.

In conclusion, the flipping process involves all the hallmarks of
nonequilibrium dynamics including spatial correlations, temporal
correlations, aging, and rejuvenation \cite{vp}.

\subsection{Small Segments} 

We complete the analysis with a direct solution for the state of small
segments containing the front. The $k$ leftmost sites can be in any
one of $2^{k-1}$ possible configurations. The equations describing the
configuration probabilities are hierarchical: due to the front motion,
the state of small segments containing the front is coupled with the
state of larger segments. To overcome this closure issue, we propose
an approximation where the state of the system outside the segment of
interest is completely random, as in our simulation method. Clearly,
this approximation becomes exact as $k\to\infty$.

A segment of length two can be in one of two configurations: $10$ or
$11$. The respective probabilities $P_{10}$ and $P_{11}$ evolve
according to
\begin{subequations}
\begin{align}
\frac{dP_{10}}{dt} &= -P_{10}+P_{11}+\frac{1}{2}\,P_{11}+\frac{1}{2}\,P_{10}
\label{10}\\
\frac{dP_{11}}{dt} &=
-2P_{11}+\frac{1}{2}\,P_{11}+\frac{1}{2}\,P_{10}\,. 
\label{11}
\end{align}
\end{subequations}
We explain the latter equation in detail.  The loss rate in
\eqref{11} equals two because any of the two occupied sites may
flip. If the front flips, there is advancement, and since the second
site is occupied with probability $1/2$, the gain terms are
$\frac{1}{2}P_{11}$ and $\frac{1}{2}\,P_{10}$. The steady state
solution is $(P_{10},P_{11})=\frac{1}{4}(1,3)$; therefore
$\rho_1=1/4$. We denote by $v_k$ the velocity obtained from a
segment of length $k$. For $k=2$ we have $v_2=P_{10} + 3P_{11}$
since the front advances by one site when the front flips in the
state $10$, but it advances three sites in the state $11$ (two sites
plus an average of one, given the random occupation outside the
segment).

\begin{table}
\label{v-prop}
\begin{tabular}{|l|l|l|l|l|l|}
\hline
$k$&$v_k^{(0)}$&$v_k^{(1)}$&$v_k^{(2)}$&$v_k^{(3)}$&$v_k^{(4)}$\\
\hline
2&1.500000&&&&\\
3&1.535714&1.418947&&&\\
4&1.587165&1.826205&1.779225&&\\
5&1.629503&1.773099&1.765862&1.764458&\\
6&1.662201&1.766730&1.764592&1.758245&1.762322\\
7&1.687108&1.765129&1.763533&1.770104&1.765175\\
8&1.705987&1.764330&1.762272&1.761669&\\
9&1.720251&1.763754&1.761864&&\\
10&1.730993&1.763313&&&\\
11&1.739055&&&&\\
\hline
\end{tabular}
\caption{The velocity $v$, obtained by successive iterations of the
Shanks transformation (propagating fronts).}
\end{table}

For $k=3$, the governing equations are
\begin{subequations}
\begin{align}
\frac{dP_{100}}{dt}&=-P_{100}+\frac{3}{2}P_{101}+\frac{1}{4}P_{110}+\frac{5}{4}P_{111}\\
\frac{dP_{101}}{dt}&=-\frac{3}{2}P_{101}+\frac{5}{4}P_{110}+\frac{1}{4}P_{111}\\
\frac{dP_{110}}{dt}&=\frac{1}{2}P_{100}-\frac{7}{4}P_{110}+\frac{5}{4}P_{111}\\
\frac{dP_{111}}{dt}&=\frac{1}{2}P_{100}+\frac{1}{4}P_{110}-\frac{11}{4}P_{111}.
\end{align}
\end{subequations}
The steady state solution is
$(P_{100},P_{101},P_{110},P_{111})=\frac{1}{56}(27,11,12,6)$; thus,
the densities are $\rho_1=9/28$, and $\rho_2=17/56$ and the velocity
is $v_3=43/28$. Furthermore, $v_4=10907/6872$ and the approximation
steadily improves as $k$ increases.

We can compute the configuration of segments with  $k\leq
12$ as detailed in Appendix A. To extrapolate the velocity, we use
the Shanks transformation \cite{bo}
\begin{equation}
\label{Shanks}
v_k^{(m+1)}=\frac{v_{k-1}^{(m)}v_{k+1}^{(m)}-v^{(m)}_kv^{(m)}_k}
{v^{(m)}_{k-1}+v^{(m)}_{k+1}-2v^{(m)}_k}
\end{equation}
where $v_n^{(m)}$ is the velocity estimate after $m$ iterations.
Repeated Shanks transformations give a useful estimate for the
propagation velocity (see Table II),
\begin{equation}
\label{v-Shanks}
v_{\rm shanks}=1.76\pm 0.01.
\end{equation}
The Shanks transformation can be used to estimate other quantities as
well.  For example, we obtain an excellent estimate for the density of
the first site, \hbox{$\rho_1=0.3492\pm 0.0001$}.

\section{Pinned Fronts}

As discussed above, the quasi-static approximation neglects the
movement of the front and possible correlations between sites. Of
these two assumptions, the latter is more significant. We therefore
modify the original flipping process and forbid the front from
changing state. This minor modification pins the front and allows us
to focus on the role of correlations.

In the pinned process, a flip event at every site other than the
origin changes the state of the system exactly as in
\eqref{process}, but a flip event at the origin yields
\begin{equation}
\label{modified} \sigma_i\to 1-\sigma_i, \quad{\rm for\ all}\quad
i>0.
\end{equation}
Hence, the site at the origin is always occupied, $\sigma_0(t)=1$.

Remarkably, pinning the front results in only minor quantitative
changes. All quantities of interest including the velocity $v$, the
diffusion coefficient $D$, the decay constant underlying the decay of
the segment density $\lambda$, and the density profile $\rho_k$ are
all within a few percent of the corresponding values for propagating
fronts (Table II). In particular, the discrepancy in the propagation
velocity is smaller than $1\%$,
\begin{equation}
\label{v-pinned}
v_{\rm pinned}=1.7753\pm0.0001.
\end{equation}
We note that the velocity and the diffusion coefficient are obtained
by using the running total of segment lengths at the time when the
origin causes a flip as a surrogate for the front position
$x$. Finally, we can not exclude the possibility that the string
probability $S_n$ is characterized by the same parameter $\lambda$ in
both processes (Table II).

\begin{table}
\label{compare}
\begin{tabular}{|c|c|c|}
\hline
{\rm quantity}&{\rm propagating fronts}&{\rm pinned fronts}\\
\hline
$v$&$1.7624$&$1.7753$\\
$D$&$2.856$&$3.178$\\
$\lambda$&$0.74$&$0.75$\\
\hline
$\rho_1$&$0.3492$&$1/3$\\
$\rho_2$&$0.3400$&$1/3$\\
$\rho_3$&$0.3479$&$41/120$\\
\hline

\hline
\end{tabular}
\caption{The velocity $v$, the diffusion coefficient $D$, the decay
constant governing the string probability $\lambda$, and the first
few densities $\rho_k$ for propagating and pinned fronts. The
results are from Monte Carlo simulations in a system of size $L=200$
evolved up to time $t=10^{11}$. The exact solution for the density
profile is detailed below.}
\end{table}

\begin{figure}[t]
\includegraphics[width=0.45\textwidth]{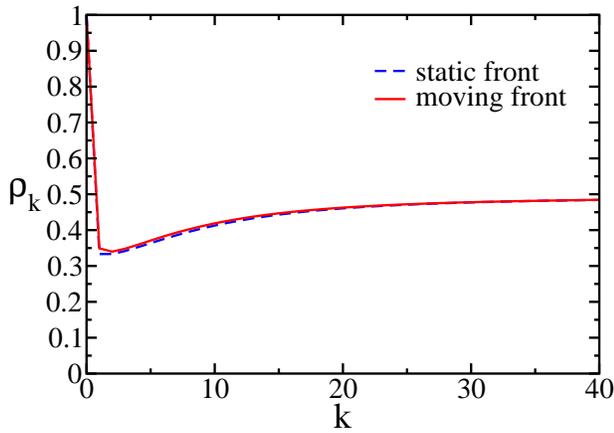}
\caption{The density $\rho_k$ versus distance $k$ for pinned and
propagating fronts. Both profiles are obtained using Monte Carlo
simulations.}
\label{fig-ck1}
\end{figure}

In addition, pinned fronts and propagating fronts have very similar
density profiles (figure \ref{fig-ck1}). The quasi-static
approximation, which is better suited for pinned fronts, becomes
slightly more accurate. Of course, the exact tail behavior
\eqref{rhok-tail} and the logarithmic excess \eqref{excess} extend
to pinned fronts.

\subsection{Correlations} 
The hierarchical evolution equation \eqref{rho-eq} for the average
occupation only assumes that the front is pinned and hence, this
equation provides an exact description. Therefore, the single-site
averages and the two-site averages are related,
\begin{equation}
\label{corr1} \langle \sigma_k\rangle - \sum_{j=0}^{k-1}\langle
\sigma_j\rangle = - 2\sum_{j=0}^{k-1}\langle \sigma_j
\sigma_k\rangle,
\end{equation}
at the steady state. Of course, $\langle \sigma_0\rangle=1$.

We can obtain the nearest-neighbor correlation $\langle
\sigma_k\sigma_{k+1}\rangle$, a quantity that evolves according to
\begin{eqnarray}
\label{nn-eq} \frac{d\langle \sigma_k\sigma_{k+1}\rangle}{dt}=
&-&\left\langle \left(2+\sum_{j=0}^{k-1}
\sigma_j\right)\sigma_k\sigma_{k+1}\right\rangle\\ +\langle
(1-\sigma_k)\sigma_{k+1}\rangle&+& \left\langle
\left(\sum_{j=0}^{k-1}\sigma_j\right)(1-\sigma_k)(1-\sigma_{k+1})\right\rangle.\nonumber
\end{eqnarray}
This equation is very similar to the equation governing the one-site
correlation. The rate of change for two occupied sites is
$2+\sum_{j=0}^{k-1}\sigma_j$ because either one of the two sites can
flip. In general, the equation for two-site correlations involves
three-site correlations, but in the particular case of neighboring
sites, the three-site correlation cancels in \eqref{nn-eq}!  We
therefore obtain a relation between average densities and two-site
correlations
\begin{equation}
\label{corr2} \sum_{j=0}^{k-2}\langle \sigma_j\rangle = \langle
\sigma_{k-1}\sigma_k\rangle +\sum_{j=0}^{k-2}\langle \sigma_j
\sigma_{k-1}\rangle +\sum_{j=0}^{k-1}\langle \sigma_j
\sigma_k\rangle.
\end{equation}
There are two different relations between the average density and
the two-site correlation: equations \eqref{corr1} and \eqref{corr2}.
By manipulating the two, we obtain the nearest-neighbor correlation
in terms of the average density,
\begin{equation}
\label{corr12} \langle \sigma_k \sigma_{k+1}\rangle =
\frac{1}{2}\,\langle \sigma_{k+1}\rangle
\end{equation}
for $k>0$. This relation demonstrates that neighboring sites are
positively correlated,
\begin{equation}
\label{corr12a} \langle \sigma_k \sigma_{k+1}\rangle -\langle
\sigma_k\rangle \langle \sigma_{k+1}\rangle=
\bigg(\frac{1}{2}-\langle \sigma_k\rangle\bigg)\,\langle
\sigma_{k+1}\rangle.
\end{equation}
We also note that correlations decay slowly at large distances as
equations \eqref{rhok-tail} and \eqref{corr12a} imply \hbox{$\langle
\sigma_k \sigma_{k+1}\rangle -\langle \sigma_k\rangle \langle
\sigma_{k+1}\rangle \simeq (4k)^{-1}$}.

For completeness, we mention that the correlation between three
consecutive sites can also be written as a function of lower-order
correlations
\begin{equation}
\langle\sigma_k \sigma_{k+1}\sigma_{k+2}\rangle=
\frac{1}{2}\langle\sigma_{k+2}(1-\sigma_k)\rangle .
\end{equation}

\subsection{Small Systems}
When the front is pinned, the system reaches a stationary state. This
steady state can be obtained exactly for small system by considering
the evolution of all possible configurations. For pinned fronts,
finite segments are not affected by flipping outside the segment, and
consequently, the evolution equations are now closed.

Consider for example a system with two sites. There are two possible
configurations: $10$ and $11$ with the respective probabilities
$P_{11}$ and $P_{10}$. These probabilities evolve according to
\begin{subequations}
\begin{align}
\frac{dP_{10}}{dt}&=-P_{10}+2P_{11}\\
\frac{dP_{11}}{dt}&=-2P_{11}+P_{10}.
\end{align}
\end{subequations}
Hence, at the steady state, $(P_{10},P_{11})=\frac{1}{3}(2,1)$ and
consequently, $\rho_1=S_2=1/3$.

\begin{table}
\begin{tabular}{|l|l|l|l|}
\hline
$k$&$\rho_k$&$S_{k+1}$&$v_{k+1}$\\
\hline
0&$1$&$1$&$1$\\
1&$\frac{1}{3}$&$\frac{1}{3}$&$\frac{4}{3}$\\
2&$\frac{1}{3}$&$\frac{1}{6}$&$\frac{3}{2}$\\
3&$\frac{41}{120}$&$\frac{23}{240}$&$\frac{383}{240}$\\
4&$\frac{76121}{216000}$&$\frac{25577}{432000}$&$\frac{714977}{432000}$\\
\hline

\hline
\end{tabular}
\caption{The density $\rho_k$, the string density $S_k$, and the
velocity $v_k=\sum_{n=1}^k P_n$, obtained by direct solution of the
microscopic evolution equations.} 
\label{smallk}
\end{table}

Next we consider the first three sites with the four configurations
$100$, $101$, $110$, $111$. The evolution equations for the
respective probabilities are
\begin{subequations}
\label{p-2-eq}
\begin{align}
\frac{dP_{100}}{dt}&=-P_{100}+P_{110}+P_{111}+P_{101}\\
\frac{dP_{101}}{dt}&=-2P_{101}+2P_{110}\\
\frac{dP_{110}}{dt}&=-2P_{110}+P_{101}+P_{111}\\
\frac{dP_{111}}{dt}&=-3P_{111}+P_{100}.
\end{align}
\end{subequations}
The steady state solution is
$(P_{100},P_{101},P_{110},P_{111})=\frac{1}{6}(3,1,1,1)$. Therefore
$\rho_1=\rho_2=1/3$ and $S_3=1/6$. Results for $k\leq 4$ are
summarized in table \ref{smallk}.

\begin{table}[t]
\begin{tabular}{|l|l|l|l|l|l|}
\hline
$k$&$v_k^{(0)}$&$v_k^{(1)}$&$v_k^{(2)}$&$v_k^{(3)}$&$v_k^{(4)}$\\
\hline
1&1.&&&&\\
2&1.333333&1.666666&&&\\
3&1.5&1.72549&1.769737&&\\
4&1.595833&1.750742&1.773156&1.775020&\\
5&1.655039&1.762616&1.774362&1.775178&1.775278\\
6&1.693228&1.768521&1.774849&1.775239&1.775289\\
7&1.718565&1.771576&1.775065&1.775267&1.775293\\
8&1.735709&1.773205&1.775170&1.775280&1.775293\\
9&1.747473&1.774095&1.775223&1.775287&\\
10&1.755632&1.774593&1.775252&&\\
11&1.761337&1.774876&&&\\
12&1.765350&&&&\\
\hline

\hline
\end{tabular}
\caption{Iterated Shanks transformations for the velocity. The zeroth
column is from the small system solution (pinned fronts).}
\label{v-pin}
\end{table}

In general, there are $2^{k-1}$ microscopic configurations in a
system of size $k$. We can compute the stationary probabilities for
systems of size $k\leq 12$ as detailed in Appendix B. Knowledge of
these steady state probabilities yields the density $\rho_k$, the
string probability $S_k$, and hence, an estimate for the velocity
$v_k=\sum_{n=1}^k S_n$.

The velocity, as well as other quantities of interest, can be
obtained very accurately using the Shanks transformation. We find
$v_{\rm shanks}=1.7753\pm 0.0001$, in perfect agreement with the
Monte Carlo simulations \eqref{v-pinned} as shown in Table
\ref{v-pin}.

\subsection{Aging and Rejuvenation} 

\begin{figure}[t]
\includegraphics[width=0.45\textwidth]{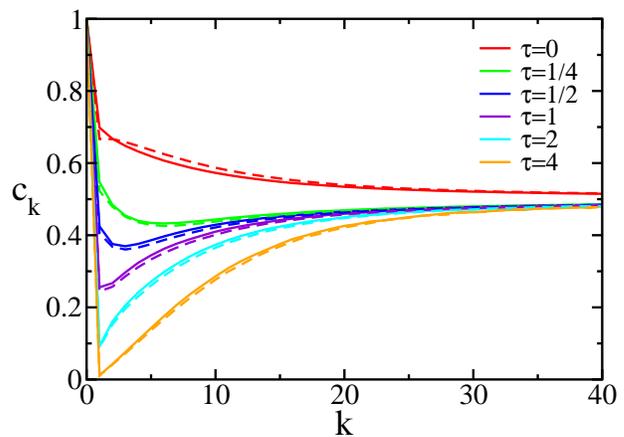}
\caption{The density $c_k$ at different ages for pinned fronts
(broken lines) and propagating fronts (solid lines).}
\label{fig-ckt-compare}
\end{figure}

We also examined the evolution with age and found that pinned and
propagating fronts display very similar behaviors, as evident from the
age-dependent density $c_k(\tau)$ (figure \ref{fig-ckt-compare}).

For pinned fronts, the zero age configuration is the exact mirror
image of the configuration just before the flip and since the front
flips at random,
\begin{equation}
\label{initial} c_k(\tau=0)=1-\rho_k
\end{equation}
for all sites except the origin, $k>0$. This expression demonstrates
the enhancement of occupied sites for newly born configurations.

Aging can be conveniently studied using small systems. For the first
site, we have $dc_1/d\tau=-c_1$ and therefore,
$c_1(\tau)=c_1(0)e^{-\tau}$. The initial condition,
\hbox{$c_1(0)=1-\rho_1=2/3$} follows from \eqref{initial}. Therefore,
\begin{equation}
c_1(\tau)=\frac{2}{3}e^{-\tau}.
\end{equation}

For the first two sites, there are four configurations: $100$,
$101$, $110$, $111$, and the respective probabilities evolve
according to
\begin{subequations}
\label{p-2-eq-a}
\begin{align}
\frac{dP_{100}}{d\tau}&=P_{111}+P_{101}\\
\frac{dP_{101}}{d\tau}&=-P_{101}+P_{110}\\
\frac{dP_{110}}{d\tau}&=-P_{110}+P_{111}\\
\frac{dP_{111}}{d\tau}&=-2P_{111}.
\end{align}
\end{subequations}
These equations differ from \eqref{p-2-eq} in that flipping events
caused by the front are excluded. The initial condition again
mirrors the stationary state
$(P_{100},P_{101},P_{110},P_{111})\big|_{\tau=0}=\frac{1}{6}(1,1,1,3)$.
By solving the evolution equations, the age-dependent density of the
second site, $c_2=P_{101}+P_{111}$, is
\begin{equation}
c_2(\tau)=\frac{1}{3}(2\tau-1)e^{-\tau}+e^{-2\tau}.
\end{equation}
Already, we can justify the non-monotonic behavior seen in figures
\eqref{fig-ckt} and \eqref{fig-ckt-compare}: $c_1>c_2$ for
$\tau<\tau_*$ with $\tau_*=0.8742$ while $c_1<c_2$ for otherwise. In
general, all densities exhibit a simple exponential decay with age,
$c_k(\tau)\propto e^{-\tau}$ as $\tau\to\infty$. We conclude that
pinned fronts faithfully capture aging and rejuvenation.

\section{Conclusions}

In conclusion, we reformulated the bit flipping process underlying the
simplex algorithm as a nonequilibrium dynamics problem and studied
spatial and temporal properties using theoretical and computational
methods. Overall, we find that the infinite interaction range leads to
rich phenomenology. There is a front that propagates ballistically
with a nontrivial velocity that is governed by the length of the
occupied strings containing the front. The propagating front includes
a deep depletion zone: vacant sites outnumber occupied sites with the
total excess of unoccupied sites growing logarithmically with
depth. The flipping process is characterized by significant spatial
correlations. For example, the likelihood of finding strings of
consecutively occupied sites is strongly enhanced.

The flipping process also exhibits nontrivial dynamics. Successive
front jumps are correlated and additionally, there are aging and
rejuvenation as young fronts are fast but old fronts are slow.
Underlying this behavior is the fact that the state of the system
just after a jump mirrors the state of the system just before a
jump.

We slightly modified the original flipping process by pinning the
front. Qualitatively and quantitatively, pinned fronts and
propagating fronts are very close. We demonstrated analytically much
of the interesting phenomenology including spatial correlations,
aging, and rejuvenation for pinned fronts.

Aging is usually characterized by using two different times
\cite{ckp}.  Here, in contrast, the time elapsed since the latest
front yields a natural definition of age and a characterization of the
dynamics that complements time itself.

We comment that there is an alternative way of studying the density
profile through an average {\it at a given lattice site} over all
realizations \cite{rbd}. The corresponding average density $\tilde
\rho_k(t)$ reaches a stationary form once the average and the
variance are taken into account,
\begin{equation}
\tilde \rho_k(t)\to \Phi\left(\frac{k-vt}{\sqrt{Dt}}\right)
\end{equation}
with $\Phi(-\infty)=0$ and $\Phi(\infty)=1/2$. This approach has a
disadvantage: the scaling function $\Phi(x)$ is dominated by
fluctuations in the position of the front. In other words, the density
profile $\rho_k$ is smeared because of diffusion.  These less
interesting diffusive fluctuations are suppressed when the front
profile is probed in a reference frame moving with the front.

We also presented a systematic solution method of small systems and
successfully demonstrated how to extrapolate relevant parameters for
infinite systems. Yet, since the complexity grows exponentially with
system size, such computations quickly become prohibitive. We have
also seen how most quantities of interest require an infinite
hierarchy of equations.  Finding an appropriate theoretical framework
with closed evolution equations remains a formidable
challenge. Nevertheless, the pinned front process provides a powerful
theoretical framework.

\acknowledgements

We are grateful for financial support from NIH grant R01GM078986, DOE
grant DE-AC52-06NA25396, NSF grants CHE-0532969 and PHY-0555312; we
also thank Jeffrey Epstein for support of the Program for Evolutionary
Dynamics at Harvard University.

\appendix

\section{Transition matrix for propagating fronts}
\label{matrix}

The evolution equations for the configuration probabilities in a
finite segment of size $k$ can be represented in the matrix form
\begin{equation}
\label{MP}
\frac{d{\bf P}}{dt} =  \mathcal{M}  {\bf P}.
\end{equation}
Here, ${\bf P}$ is the vector ${\bf P}=\{P_{1j}\,| 0\leq j\leq
U-1$\} where $j$, written as a binary, is in increasing order and
$U=2^{k-1}$. For example, when $k=5$ the state vector is
$(P_{10000}, P_{10001}, \cdots, P_{11111} )$, with $U=16$ entries.
The elements of this vector equal the probabilities that the system
is in the respective configuration.  Also, $\mathcal{M}$ is the
$U\times U$ transition matrix whose elements equal the transition
rates between
the corresponding configurations. 

The transition matrix $\mathcal{M}$ is a sum of three matrices
\begin{equation}
\label{emm} \mathcal{M} = \mathcal{M}_1+\mathcal{M}_2+\mathcal{M}_3.
\end{equation}
We quote the first two for $k=5$,
\begin{equation*}
\label{integer}
\mathcal{M}_1 = \left( \begin{array}{rrrrrrrrrrrrrrrr}
0 & 1 && 1 &&&& 1&&&& &&&&1\\
 &  & 1&&&& 1 &&&&&&&& 1&\\
 &&& 1 && 1&&&&&&&& 1& &\\
&&&&1 &&& && &&& 1&&&\\
&&&&&1&&1&&&&1&&&&\\
&&&&&&1&&&&1&&&&&\\
&&&&&&&1&&1&&&&&&\\
&&&&&&&&1&&&&&&&\\
&&& &&&&&&1&&1&&&&1\\
&&&&&&&&&&1&&&&1&\\
&&&&&&&&&&&1&&1&&\\
&&&&&&&&&&&&1&&&\\
&&&&&&&&&&&&&1&&1\\
&&&&&&&&&&&&&&1&\\
&&&&&&&&&&&&&&&1\\
 &&&&&&&&&&&&&&&0
 \end{array} \right)
\end{equation*}
and
\begin{equation*}
\label{fractional}
\mathcal{M}_2 = \frac{1}{16}\times \left( \begin{array}{rrrrrrrrrrrrrrrr}
&&&&&&&8&&&&4&&2&1&1\\
&&&&&&&8&&&&4&&2&1&1\\
&&&&&&8&&&&&4&&2&1&1\\
&&&&&&8&&&&&4&&2&1&1\\
&&&&&8&&&&&4&&&2&1&1\\
&&&&&8&&&&&4&&&2&1&1\\
&&&&8&&&&&&4&&&2&1&1\\
&&&&8&&&&&&4&&&2&1&1\\
&&&8&&&&&&4&&&2&&1&1\\
&&&8&&&&&&4&&&2&&1&1\\
&&8&&&&&&&4&&&2&&1&1\\
&&8&&&&&&&4&&&2&&1&1\\
&8&&&&&&&4&&&&2&&1&1\\
&8&&&&&&&4&&&&2&&1&1\\
8&&&&&&&&4&&&&2&&1&1\\
8&&&&&&&&4&&&&2&&1&1
 \end{array} \right)
\end{equation*}
The third matrix $\mathcal{M}_3$ is diagonal and it guarantees that
each column of $\mathcal{M}$ sums to zero. We note that the transition
matrix is sparse.  The steady state probability equals the zeroth
eigenvector, $\mathcal{M} {\bf P}=0$. Finally, the velocity follows
from the average advancement expected in each configuration. This
advancement is represented by the vector ${\bf J}$ and for example,
${\bf J}=(1,1,1,1,1,1,1,1,2,2,2,2,3,3,4,6)$ for $k=5$. The velocity is
simply the scalar product, $v={\bf J}\cdot{\bf P}$.

\section{Transition matrix for pinned fronts}

Using the matrix notation in \eqref{MP}, the evolution equations for
$k=4$ involve the following transition matrix
\begin{equation}
\label{M4} \mathcal{M} = \left(
\begin{array}{rrrrrrrr}
-1&1&&1&&&&2\\
\\
&-2&1&&&&2&\\
\\
&&-2&1&&2&&\\
\\
&&&-3&2&&&\\
\\
&&&1&-2&1&&1\\
\\
&&1&&&-3&1&\\
\\
&1&&&&&-3&1\\
\\
1&&&&&&&-4\\
\end{array} \right).
\end{equation}
In this case, the steady state probabilities are
\begin{equation}
\label{P4-sol}
{\bf P} = \left( \begin{array}{c}
P_{1000} \\
P_{1001} \\
P_{1010}\\
P_{1011}\\
P_{1100} \\
P_{1101} \\
P_{1110}\\
P_{1111}
\end{array} \right) =
\frac{1}{240}\times \left( \begin{array}{c}
92\\
28\\
22\\
18\\
27\\
13\\
17\\
23
\end{array} \right)
\end{equation}
Thus, the density is $\rho_3=41/120$ and the string probability is
$P_4=23/240$.


\begin{thebibliography}{99}
\bibitem{note} It is simple to show for a renewal process,
     \hbox{$D_{\rm renew}=\frac{1}{2}\langle n^2\rangle$}, but this
     value underestimates the diffusion coefficient, \hbox{$D_{\rm
     renew}=2.750\pm 0.001$}.
\end{thebibliography}
\end{document}